\documentstyle[aps]{revtex}
\def\be{\begin{equation}}
\def\ee{\end{equation}}
\def\al{\alpha}
\def\bt{\beta}
\def\ld{\lambda}
\def\sg{\sigma}
\def\eps{\varepsilon}
\def\th{\vartheta}
\def\Re{{\rm Re}}
\def\Im{{\rm Im}}

\begin{document}
\twocolumn

\title{On the Muon Decay Parameters.}

\author{M. V. Chizhov}

\address{
PPE Division, CERN, CH-1211 Geneva 23, Switzerland\\
and\\
Centre for Space Research and Technologies, Faculty of Physics,
University of Sofia,\\ 
1164 Sofia, Bulgaria, E-mail: mih@phys.uni-sofia.bg}

\maketitle

\begin{abstract}
Predictions for the muon decay spectrum are usually derived from the
derivative-free Hamiltonian. However, it is not the most general form 
of the possible interactions. Additional simple terms with derivatives 
can be introduced. In this work the distortion of the standard energy
and angular distribution of the electrons in polarized muon decay
caused by these terms is presented.\\
\end{abstract}


    More than sixty years have passed since the time of brilliant
Pauli's idea about neutrino~\cite{Pauli} 
and celebrated Fermi theory of $\beta$-decay~\cite{Fermi}.
Still the famous Fermi interaction

\be
G_F \left[\bar{\Psi}_p \gamma_\ld \Psi_n\right] 
\left[\bar{\Psi}_e \gamma_\ld \gamma^5 \Psi_\nu\right]
\label{Fermi}
\ee

\noindent is the corner-stone of all low energy weak processes. 
In order to investigate a more general form of the interactions,
for the particular case of the muon decay, additional terms have been
introduced by Michel~\cite{Michel}. They can be
expressed by fields of definite chiralities~\cite{Scheck} 

\begin{equation}
{\cal H}={4 G_F \over \sqrt 2}
\sum_{\mbox{\tiny
 $\begin{array}{c} k=S,V,T\\ \eps,\chi=R,L\end{array}$}}
\Bigl\{ g^k_{\eps\chi}
 \left[ \bar{e}_{\eps} \Gamma^k \nu^e_n \right]
 \left[ \bar{\nu}^{\mu}_m \Gamma^k \mu_{\chi} \right] + {\rm h.c.}
\Bigr\} . 
\label{Ham}
\end{equation}

\noindent Here, $k$ labels the type of interaction (scalar, vector,
tensor), $\epsilon$ and $\chi$ indicate the chirality of the charged
leptons. (The chiralities of the neutrinos, $n$ and $m$, are uniquely
fixed by $\eps$, $\chi$, and $k$).
These interactions differ from (\ref{Fermi}) only by the fact that we have
introduced 10 coupling constants instead of the unique one, $G_F$. 
The standard $V$-$A$ interaction implies that $g^V_{LL}=1$, and other
$g^k_{\eps\chi}$'s are zero. Nonstandard couplings may arise in the
extensions of the standard model due to the exchange of new intermediate
bosons, other than $W^\pm$. In this case the energy and angular
distribution of the emitted charged leptons depends on bilinear
combinations of the coupling constants $g^k_{\eps\chi}$ called the
Michel parameters.

It is widely accepted that this is the most general form of the
energy spectrum. In all present experiments only such form of the
energy distribution is tested. However, it is easy to see that
sensitivity to new physics for the introduced terms is feeble. In the
case of massless neutrinos only the single scalar interaction with the
coupling constant $g^S_{RR}$ interferes with the standard $V$-$A$
interaction and it can give considerable contribution to the muon
decay spectrum. This is parametrized by the Michel parameter $\eta$.
Unfortunately, when the polarization of the final charged leptons is
not detected, the sensitivity to $\eta$ is diminished by the small
ratio of the mass of the final charged lepton to its maximal energy
$x_0=m_e/E_{max}$.  To determine the parameter $\eta$ without a
suppression factor the transverse polarization of the produced
charged lepton has been measured, but with the smaller accuracy than
the Michel parameters $\rho$, $\delta$, and $\xi$. The latter
parameters are influenced by other nonstandard terms in (\ref{Ham}),
which do not interfere with the standard $V$-$A$ interaction and the
sensitivity to a new physics from the experimental bounds on the
nonstandard coupling constants is negligible.

      It is possible to introduce new interactions of a
simple form, which would interfere with the primary $V$-$A$
interaction without suppression. To do this we assume that the
effective Hamiltonian depends on the momentum transfer $q_\ld$ or
the derivative $\partial_\ld$ of the charged current. Then the
simplest form of the new interactions is

\begin{eqnarray}
{\cal H}_T &=& {4G_F\over \sqrt 2 m_\mu}\left\{
f_\mu~ [\bar{e}_L\gamma_\al \nu^e_L]~
\partial_\bt [\bar{\nu}^\mu_L \sg_{\al\bt} \mu_R]
\right.
\nonumber \\
&&+
\left.
f^*_e~ \partial_\bt[\bar{e}_R \sg_{\al\bt} \nu^e_L]~
[\bar{\nu}^\mu_L\gamma_\al \mu_L] + h.c.\right\}.
\label{new}
\end{eqnarray}

\noindent Such interactions can arise, for example, due to an
anomalous $W$-boson coupling with the fermions. Various anomalous
couplings of the gauge bosons are tested now at high energies.
Generally speaking the effective coupling constants of the new
interactions (\ref{new}) may depend on the momentum transfer $f(q^2)$
and such kind of effective interactions should apply only for a narrow
region of momenta. In our case we restrict ourself to low energy processes
of lepton decays and assume that the coupling constant $f$ is a constant.

The interactions (\ref{new}) lead to
the specific energy distribution, which can be tested experimentally.
This is not a new idea and analogous early attempt, made by
Konopinski and Uhlenbeck~\cite{der}, had the aim to introduce the
derivative coupling for the neutrino field. The additional
interactions (\ref{new}) include the conserved tensor currents
$J^\pm_{\al}=\partial_\bt [\bar{\Psi} \sg_{\al\bt} (1\pm\gamma^5) \Psi]$
and naturally extend the helicity projection form of the interaction 
(\ref{Ham}).
It is impossible to construct other nontrivial terms depending on the
momentum transfer, because the derivative of the vector current
$i\partial_\ld [\bar{\Psi}_\nu\gamma_\ld \Psi_l]=
m_l \bar{\Psi}_\nu \Psi_l$ reduces to the scalar term
due to the Dirac equation. Therefore, besides the trivial dependence of
the coupling constants on the momentum transfer $g(q^2)$, other
additional terms do not exist here.\footnote{We have dropped the derivative
tensor-tensor interaction 
$f_\mu f^*_e~ \partial_\al [\bar{e}_R \sg_{\al\ld}\nu^e_L]~
\partial_\bt [\bar{\nu}^\mu_L \sg_{\bt\ld} \mu_R]$ 
which includes two new currents and is negligible.}

In the following we give the prediction for the energy and angular
distribution of electrons from the decay of polarized muons
accounting for the effect of the new terms (\ref{new}). 
The generalization to leptonic $\tau$-decay is obvious.
The differential decay probability to obtain an $e^\pm$ with reduced energy
$x=E_e/E_{max}$, emitted at an angle $\th$ with respect to the muon's
polarization $\bbox{\cal P}_\mu$, and having its spin pointing into the
direction of the arbitrary unit vector $\bbox{\zeta}$ is given by~\cite{PDG}

\begin{eqnarray}
{{\rm d}^2 \Gamma \over {\rm d}x {\rm d}\cos\th} &=&
{G_F^2 m_\mu \over 12 \pi^3} E_{max}^4 \sqrt{x^2-x_0^2}
\nonumber \\
&&\times \left[F_{IS}(x)\pm {\cal P}_\mu \cos\th~ F_{AS}(x)\right]
\nonumber \\
&&\times \left[1+\bbox{\cal P}_e(x,\th) \cdot \bbox{\zeta}\right],
\end{eqnarray}

\noindent where we have used ${\cal P}_\mu=|\bbox{\cal P}_\mu|$, and
where $\bbox{\cal P}_e$ is the polarization vector of the $e^\pm$:

\be
\bbox{\cal P}_e=P_{T_1}~\bbox{x} + P_{T_2}~\bbox{y} + P_L~\bbox{z}
\ee

\noindent with components
\begin{eqnarray}
P_{T_1}(x,\th) &=& {{\cal P}_\mu \sin\th ~ F_{T_1}(x)
\over F_{IS}(x)\pm {\cal P}_\mu \cos\th ~ F_{AS}(x)}, \\
P_{T_2}(x,\th) &=& {{\cal P}_\mu \sin\th ~ F_{T_2}(x)
\over F_{IS}(x)\pm {\cal P}_\mu \cos\th ~ F_{AS}(x)}, \\
P_L~(x,\th) &=& {\pm F_{IP}(x) + {\cal P}_\mu \cos\th~ F_{AP}(x)
\over F_{IS}(x)\pm {\cal P}_\mu \cos\th ~ F_{AS}(x)}
\end{eqnarray}
 
\noindent  defined in a right-handed coordinate system:

\be
\bbox{z}={\bbox{k}_e \over |\bbox{k}_e|},~~~~~~
\bbox{y}={\bbox{k}_e \times \bbox{\cal P}_\mu \over 
|\bbox{k}_e \times \bbox{\cal P}_\mu|},~~~~~~
\bbox{x}=\bbox{y}\times\bbox{z}.
\ee

The functions $F(x)$ can be decomposed as

\be
F(x)=F^{V-A}(x) + G(x),
\ee

\noindent where the first term corresponds to the standard $V$-$A$
interaction and the second one includes leading contributions from
the interference with the new terms (\ref{new}). The functions 
$F^{V-A}(x)$ are given by 

\begin{eqnarray}
F^{V-A}_{IS}(x)&=&-2x^2+3x-x_0^2, \\
F^{V-A}_{AS}(x)&=&\sqrt{x^2-x_0^2}~(2x-2+\sqrt{1-x_0^2}), \\
F^{V-A}_{IP}(x)&=&\sqrt{x^2-x_0^2}~(-2x+2+\sqrt{1-x_0^2}), \\
F^{V-A}_{AP}(x)&=&2x^2-x-x_0^2, \\
F^{V-A}_{T_1}(x)&=&x_0(x-1), \\
F^{V-A}_{T_2}(x)&=&0.
\end{eqnarray}

\noindent Since the new functions $G(x)$ contribute to the
muon decay spectrum with a small constant $f$ we have kept only the
leading terms of the small mass parameter $x_0$:

\begin{eqnarray}
G_{IS}(x)&=&2x^2\Re f_\mu+x_0(x^2-3x+3)\Re f_e, \\
G_{AS}(x)&=&2x^2\Re f_\mu+x_0 (-x^2+2x)\Re f_e, \\
G_{IP}(x)&=&2x^2\Re f_\mu+x_0(-x^2+2x)\Re f_e, \\
G_{AP}(x)&=&2x^2\Re f_\mu+x_0(x^2+x-1)\Re f_e, \\
G_{T_1}(x)&=&(2x^2-x)\Re f_e + \frac{1}{2}x_0(2x^2-x)\Re f_\mu, 
\label{T1} \\
G_{T_2}(x)&=&(2x^2-x)\Im f_e + \frac{1}{2}x_0(2x^2-3x)\Im f_\mu
\label{T2}.
\end{eqnarray}

Both terms in (\ref{new}) contribute to the energy distribution of the
polarized electrons from the muon decay without suppression.
Therefore, experimental bounds on the new energy dependences directly
constrain the strength of the new interactions. Such kind of the 
spectrum has not been tested experimentally. In order to estimate
the strength of the new interactions we shall use the experimental
values of the Michel parameters. Let us consider the muon
decay, whereas the errors on the Michel parameters are typically at
the per-mill level. The most precisely known is the $\rho$ parameter
$\rho=0.7518\pm0.0026$. This puts a strong constraint on the new parameter

\be
\Re f_\mu < 4.5 \times 10^{-3}.
\label{f}
\ee

\noindent Contrary to the $\rho$ parameter the new parameter $f$
contributes to the muon decay width

\be
\Gamma={G_F^2 m_\mu^5 \over 192 \pi^3} (1+f_\mu)
\ee

\noindent and influences the value of the Fermi coupling
constant $G_F$. This constant is used in all precision experiments in
electroweak physics and it is obvious that its value cannot
differ significantly from the standard magnitude 
$G_F^{SM}=1.16639(2) \times 10^{-5}$ GeV$^{-2}$. A self-consistent check
gives the unitarity of the Cabibbo-Kobayashi-Maskawa matrix. Our
present knowledge of the matrix elements comes from the comparison of
the strength of the quark weak interaction to the muon decay~\cite{PDG}.
Therefore, effectively we can write

\be
|V_{ud}|^2+|V_{us}|^2+|V_{ub}|^2=
\left({G_F \over G_F^{SM}}\right)^2=0.9965\pm0.0021
\ee

\noindent that leads to similar bound on the parameter 
$\Re f_\mu < 5.6 \times 10^{-3}$ as in Eq.\ (\ref{f}).

The measurement of the electron decay asymmetry

\be
{\cal A}(x)={\cal P}_\mu {F_{AS}(x) \over F_{IS}(x)}
\ee

\noindent from polarized muons determines the Michel parameters
$\delta=0.749\pm0.004$ and ${\cal P}_\mu \xi=1.003\pm0.008$.
They lead also to the analogous constraints 
$\Re f_\mu(\delta) < 7 \times 10^{-3}$ and 
$\Re f_\mu(\xi) < 5.5 \times 10^{-3}$, respectively.
It is interesting to note that in the special case 
${\cal A}(1)={\cal P}_\mu~\xi \delta/ \rho$ at the spectrum end point
our prediction coincides with the standard model 
${\cal A}^{SM}(1)={\cal P}_\mu$.

The measurement of the longitudinal polarization $P_L$ of the
electron from the decay of unpolarized and polarized muons allows one
to determine the Michel parameters $\xi'$ and $\xi''$.
The new interaction (\ref{new}) does not effect the $\xi'$ parameter and
no constraint on the $f$ parameter can be imposed. The present precision
of the measured combination~\cite{L} $\xi''/\xi-\xi'=-0.35\pm0.33$
does not lead to better limit on the $f$.

The transverse electron polarizations $P_{T_1}$ and $P_{T_2}$
allow to constrain the $\eta$ parameter and a $CP$-violation
phase. In our case it will be a limit on the $f_e$ parameter. Although
the accuracy of the $P_{T_1}$ determination is high~\cite{T}
$\langle P_{T_1} \rangle=0.016\pm0.023$
it can not constrain very well the new parameter

\be
\Re f_e < 7.4 \times 10^{-2},
\ee

\noindent like $\eta=-0.007\pm0.013$ due to the lower sensitivity of the
$P_{T_1}$ to the $f_e$. The same things concern the determination
of the $\Im f_e$ or the $CP$-violation phase.

Concerning the present experimental situation and future plans we should 
mention the following experiments. The planed experiment~\cite{exp} 
in Villigen PSI-R-94-10 can improve by an
order of magnitude the previous measurement of the transverse
polarization components of the positron and can provide a tighter
limit for the new parameter $f_e$ as well. It is very interesting
also to fit the new energy and angular positron distribution from the
polarized $\mu^+$ in the completed Los Alamos LAMPF-1240 experiment 
and take it into account in the future TRIUMF-614 experiment. 
The goal of the E614 Collaboration is the measurement of the 
parameters $\rho$, $\xi$, and $\delta$ to precisions of few $10^{-4}$!

What concerns the $\tau$-decays, assuming universality of the new
interactions, the statistics collected at the LEP will allow to
extract the $f_\tau$ parameter with the precision better than $10^{-3}$.
This can be done similarly to the polarization study for semileptonic and
pure leptonic $\tau$-decays performed by the LEP Collaborations.

In conclusion the author is glad to thank M. Shaposhnikov, E. Fradkin, 
L. Alvarez-Gaum\'{e}, D. Bardin, L. Litov, A. Olchevsky, I. Boyko 
and D. Kirilova for useful discussions and encouragements.
The author acknowledge the warm hospitality of the DELPHI
Collaboration at CERN, where this work has been fulfilled.

\vspace{1cm}

\pagebreak[1]

\end{document}